\documentclass[prb,twocolumn,showpacs,floatfix]{revtex4}
\usepackage{pdfsync}
\usepackage{times}
\usepackage{bm}
\usepackage[dvips]{graphicx}
\usepackage{amsmath}
\usepackage{natbib}
\usepackage{color}
\usepackage{gensymb}
\usepackage{cleveref}

\begin{document}

\title{Velocity renormalization and Dirac cone multiplication in graphene 
superlattices with various barrier edge geometries}

\author{A.~de~Jamblinne~de~Meux$^{1}$, N.~Leconte$^{1,2}$, J.-C.~Charlier$^{1}$, 
 A.~Lherbier$^{1}$}

\affiliation{
$^1$Institute of Condensed Matter and Nanosciences (IMCN), Universit\'e catholique de Louvain (UCL), Chemin des \'etoiles 8, B-1348 Louvain-la-Neuve, Belgium\\
$^2$ICN2 - Institut Catala de Nanociencia i Nanotecnologia, Campus UAB, 08193 
Bellaterra (Barcelona), Spain\\
}

\date{\today}

\begin{abstract} 

The electronic properties of one-dimensional graphene superlattices strongly 
depend on the atomic size and orientation of the 1D external periodic potential. 
Using a tight-binding approach, we show that the armchair and zigzag directions in these superlattices have a different impact on the renormalization 
of the anisotropic velocity of the charge carriers. For symmetric potential 
barriers, the velocity perpendicular to the barrier is modified for the armchair direction while remaining unchanged in the zigzag case. For asymmetric barriers, the initial symmetry between the forward
and backward momentum with respect to the Dirac cone symmetry is broken for the 
velocity perpendicular (armchair case) or parallel (zigzag case) to the 
barriers. At last, Dirac cone multiplication at the charge neutrality point occurs only for the zigzag geometry. In contrast, band gaps appear in 
the electronic structure of the graphene superlattice with barrier in the armchair direction.

\end{abstract} 

\pacs{72.80.Vp 73.22.Pr} 

\maketitle

  \section{Introduction} 

Graphene has attracted much attention since 2004~\cite{grapheneDisc} with the 
first experimental characterizations unveiling its amazing electronic and 
transport properties~\cite{Novoselov2011837}. The unique behavior of massless 
chiral electrons present in graphene gives rise to uncommon effects such as the 
minimum of conductivity~\cite{riseofgr}, Klein tunneling~\cite{Katsnelson2006620} 
of charge carriers and the anomalous quantum hall 
effect~\cite{novoselov2005two}. Such tremendous properties related to 
pseudo-relativistic effects have motivated intense efforts towards the 
realization of graphene-based electronics~\cite{Lemme2010, Yamaguchi2010524, 
Wu201174, Palacios2011464, Schwierz2010487}.
  
Within the variety of graphene-based devices for which a great potential is 
foreseen, the specificities of graphene superlattices have triggered special 
interest~\cite{Park,Park2,PhysRevB.83.195434,PhysRevLett.100.056807,
PhysRevLett.102.056808,PhysRevB.79.205435,guinea2010band,Pedersen2008, 
Furst2009, Giesbers2012, Eroms2009, Pedersen2014}. In particular, 
one-dimensional (1D) graphene superlattices (SLs), which can be obtained by 
controlled surface patterning, might exhibit electron beam 
supercollimation~\cite{park2008electron}.  Furthermore, the possibility of 
creating new Dirac cones at the charge neutrality point (CNP) or at low 
excitation energy is also an interesting feature of graphene SLs, both in single 
and bi-layer graphene~\cite{barbier2010single}. The appearance of new Dirac 
cones corresponds to observable dips in the density of state (DOS), which were 
observed experimentally by STM measurements on graphene exposed to a Moir\'e 
pattern~\cite{Yankowitz2012382, Ponomarenko2013594, Dean2013598, Hunt20131427, 
Yang2013792} and in corrugated graphene~\cite{corrugate}. 1D SLs can serve as 
simplified models for both corrugated graphene and graphene modulated by an 
external electrical field.

The directional dependency of the electron propagation in 1D graphene SLs is 
related to an unexpected anisotropy of the electron velocity.  In the direction 
perpendicular to the potential barrier, the velocity of the electron remains 
constant and is always equal to the velocity of pristine graphene. This behavior 
is unaffected by the width ($W$), the height ($U$) or the modulation period 
($L$) of the barrier. In sharp contrast, the velocity parallel to the barrier is 
strongly impacted by these three parameters. In certain cases, the velocity in 
the direction parallel to the barrier can even completely 
vanish~\cite{Park,Park2}. Further experimental work confirmed and extended these 
first observations. In particular, Dubey \textit{et 
al.}~[\onlinecite{TunableSuperl}] succeeded to fabricate 1D SLs and observed the 
variation of the resistance of this structure as a function of the barrier 
height. Their observations corroborate the appearance of new Dirac cones at the 
charge neutrality points predicted by Barber \textit{et 
al.}~[\onlinecite{Barbier10}] and Ho \textit{et al.}~[\onlinecite{Ho09}]. 
Including the influence of the barriers edges, Lee \textit{et 
al.}~[\onlinecite{Modificationo}] showed by \textit{ab-initio} calculation that 
gaps can appear in 1D SLs with barriers in the armchair direction, while this is 
not the case in the zigzag direction. In the experimental part, the graphene 
samples were suspended over periodic nanotrenches patterned in the substrate. 
Since interaction with the substrate is believed to introduce chemical doping 
through the adsorption of oxygen, the periodic potential was modeled by epoxy 
oxygens ad-atoms which can lead to impurity specific physics, such as resonant 
states leading to localization effects~\cite{PhysRevB.84.235420}. It is therefore not 
clear if the the apparition of those gaps are due to the oxygens ad-atoms or to 
the presence of the SL.
  
The first results by Lee \textit{et al.}~[\onlinecite{Modificationo}] and the known importance of edge physics in graphene ribbons~\cite{nakada1996edge, 
PhysRevLett.97.216803} call for a more systematic analysis, taking into account the effect of edge physics of the barriers in 1D SLs. The experimental realization of controlled 1D SLs is possible, albeit a challenging task. If the chiral directions of a graphene sheet are known, alignment of periodic gates on a patterned surface is possible. As an example, Ponomarenko et al. successfully align graphene on hexagonal boron nitride\cite{coningCone} . The knowledge of the chirality is more challenging. A first possibility is through CVD grown graphene which was reported to exhibit hexagonally shaped grains with aligned edges, mostly, in the zigzag direction\cite{yu2011control}. Another solution would be to etch a small portion of the graphene via iron catalysts\cite{lei2015water}. As the chirality along the etched line is preserved, alignment becomes possible in this direction.  
   
  In this article, the differences between the zigzag SLs (ZSLs) and armchair 
SLs (ASLs) are investigated systematically by comparing the impact of the three 
relevant parameters defining a 1D SL: $U$, $W$ and $L$. The following 
conclusions are drawn. By taking into account the different orientations of the 1D SL (zigzag and armchair), the Dirac cone multiplication behavior is complexified. The energy 
position of the additional Dirac cones depends on the value of U, as 
predicted by Park \textit{et al.}~[\onlinecite{Park}]. In addition, the new 
cones at the charge neutrality point (CNP) described by Ho \textit{et 
al.}~[\onlinecite{Ho09}] only appear for ZSLs and lead to the apparition of 
three families of cones. In contrast, for ASLs, band gaps appear in the density 
of states. Depending on the number of carbon dimers making up the barrier, ASLs 
can be classified into three families. Finally, the velocity modulation 
quantified by Barbier \textit{et al.}~[\onlinecite{Barbier10}] is also impacted 
by the SL barrier edge geometry. For symmetric barriers, the velocity perpendicular to the 
barriers is modified for ASLs and is robust in the case of ZSLs. For asymmetric 
barriers, velocity renormalization is observed in both directions for all SLs. 
These asymmetric barriers also break the initial symmetry between the forward
and backward momentum direction for one of the two main directions of the 
velocity: in the perpendicular direction for ASLs and in the parallel direction 
for ZSLs.

  This systematic analysis is organized into three different Sections in this 
paper, namely the impact of the parameters defining the barrier on the DOS and 
the band structure (Section~\ref{DOSandBandstructure}), on the velocity at the 
CNP (Section~\ref{velocityAtCNP}) and finally, on the velocity at higher 
energies (Section~\ref{velocityAtEnergies}). The numerical techniques and the 
model used to describe the SLs are presented in 
Section~\ref{numericalTechniques}.Section \ref{disorderEffectSection} describes the effect of white noise disorder on our calculations.

  \section{Numerical Techniques and Model}
  \label{numericalTechniques}

The Hamiltonian is expressed in an orthogonal single $p_z$ orbital basis set. The tight-binding model accounts only for first nearest-neighbors interactions described by the hopping term $\gamma_0$=-2.6 eV. The Hamiltonian of the two-dimensional periodic system reads as

\begin{equation}
H = \epsilon_z \sum_i c_i^\dagger c_i + \gamma_0 \sum_{<i,j>} c_i^{\dagger} c_j
\end{equation}

where $c_i^\dagger$ and $c_i$ are creator and annihilator operator on atomic site $i$ and where $<i,j>$ denotes the sum runs only on atomic site $j$ being first nearest neighbors of atomic site $i$.
The onsite energy term ($\epsilon_z$) accounts for the local electrostatic environment. 

To create a 1D periodic superlattice potential on top of the 2D graphene plane, two electrostatic regions
can be defined by setting $\epsilon_z$ to $\epsilon_1$ or $\epsilon_2$ depending on whether the atomic site belongs to the first or the second electrostatic region. No other modifications than the onsite terms of the Hamiltonian are performed to account for the presence of the electrostatic barrier potential. In particular, the hopping terms between atomic sites belonging to the two separate regions remain unchanged (i.e. equal to $\gamma_0$). While cutting graphene into nanoribbons will produce physical edges, here the system remains an infinite 2D graphene plane but with a 1D periodic potential imposing a given 1D orientation with respect to the crystal.

\begin{figure}[tb]
    \center
    \includegraphics[width=8cm]{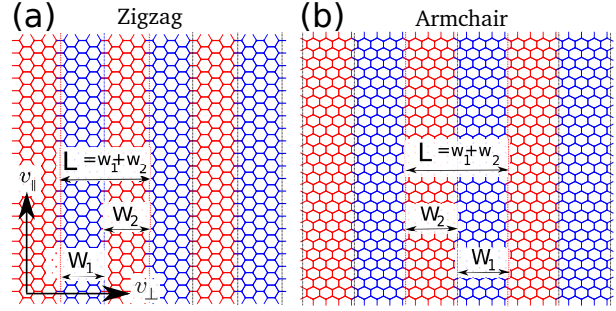} 
    \caption{Representation of graphene SLs with the potential applied in (a) 
the zigzag direction and (b) the armchair direction. The blue regions contain 
atoms with an applied potential $\epsilon_1$ that form potential barriers of 
width $W_1$. The red regions form barriers of width $W_2$ and contain atoms with 
applied potential $\epsilon_2$. $L$ is the periodicity of the pattern.}
    \label{KPP}
  \end{figure}
 
Figure~\ref{KPP} depicts how ZSLs and ASLs are modeled. Their structures 
are composed of two potential barriers of widths $W_1$ and $W_2$ repeated 
periodically with period $L=W_1+W_2$. The barrier heights are given by 
$\epsilon_1$ and $\epsilon_2$ respectively. Actually, only the potential difference $U=\epsilon_1-\epsilon_2$

induces a modification of the electronic 
properties discussed in this paper, notwithstanding a rigid shift of the CNP, 
which is given by 
  \begin{equation}
    \frac{W_1\epsilon_1+W_2\epsilon_2}{L}.  
    \label{shiftCNP}
  \end{equation}

It is therefore easier to work with an effective potential U by setting $\epsilon_1=0$ eV and $\epsilon_2= U$, and to realign a posteriori the CNP to zero energy in order to compare the different systems.
Only moderate values of $U$ are considered, as 
larger values are not achievable experimentally and because the model becomes 
unreliable for large value of $U$. 
Finally, in this article, $W=W_2$ and, unless stated otherwise, $W_2=L/2$ 
(symmetric barrier). 

  The presence of a 1D periodic potential, creating a regular superlattice, 
maintains the transport in the ballistic regime, \textit{i.e.} the propagation 
is governed by the periodic extended Bloch states. Therefore in such perfect SLs,
the only relevant quantities governing the ballistic transport are the DOS 
and the carrier velocities.
The velocity can be evaluated from electronic band structure computed 
by a direct diagonalization of the Hamiltonian around a given $k$ point:

  \begin{equation}
    v_k=\frac{1}{\hbar} \frac{\partial E(k)}{\partial k}.
    \label{vFromDiag}
  \end{equation}

The energy-dependency of the velocity $v(E)$ can then be computed by integrating over the whole Brillouin zone.
Alternatively, the real-space TB-based Kubo-Greenwood formalism, as described in 
Refs.~[\onlinecite{thesisaurelien,PhysRevLett.79.2518,PhysRevB.59.2284,PhysRevB.84.235420,
Roche20121404,fan2014efficient,radchenko2012influence,trambly2011electronic}], 
can be used to compute carrier velocities from the spreading of wavepackets in the ballistic regime.
Unfortunately for graphene, the velocity cannot be accurately 
computed at the CNP in this formalism because a mathematical singularity exists at this 
point (see appendix~\ref{instability} for more information). 
The other transport quantities such as the semi-classical conductivity, 
the mean free path or the mobility, typically accessible within the Kubo-Greenwood approach,
are only well defined into the diffusive regime. 
This regime is however only obtained after multiple scattering events in 
presence of a stochastic disorder potential.
The Kubo-Greenwood method can also describe quantum localization effects in disordered systems. 
The present paper mainly focuses on perfect systems, where charge 
carriers remain in the ballistic regime for the whole energy spectrum.
The impact of disorder is discussed in Section~\ref{disorderEffectSection} at the end of the paper.

Finally, the DOS can be calculated independently of the time evolution of wavepackets thanks to the 
Haydock recursion method using a Lanczos algorithm with continued fraction calculations~\cite{Haydock198011}.
  
  \section{Density of state and electronic bandstructure}
  \label{DOSandBandstructure}

  \subsection{Generic characteristics for ASLs and ZSLs}

\begin{figure}[h]
\center
\includegraphics[width=7cm]{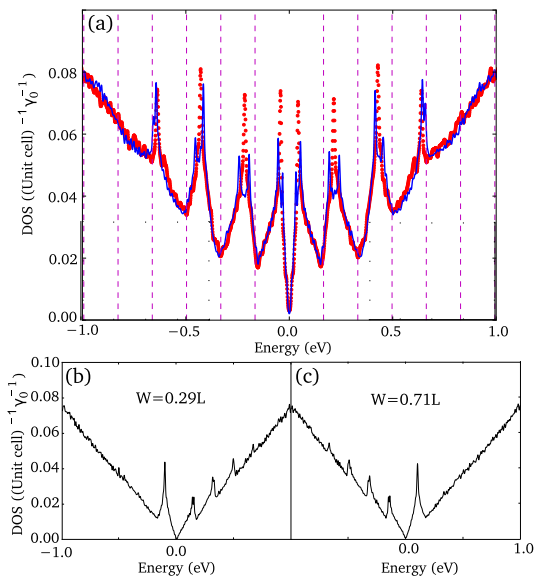} 
\caption{DOS of SLs. (a) Difference between the DOS for a ASL (dotted red 
points) and ZSL (solid blue line). Both DOS are given for a potential $U=1.04$ 
eV and a period $L$ of $10$ nm. The dashed vertical lines gives the position of 
the new Dirac cones as given by Eq.~(\ref{ParkFormula}). The only visible 
difference is the splitting of the peaks for ZSL. Antisymmetric DOS around the 
CNP for ZSLs with $W=0.29 L$ (b) and $W=0.71 L$ (c) ($L=10$ nm, $U=0.52$ eV).} 
\label{doszzarm}
\end{figure}

  The main impact of a SL on the DOS is the presence of the local dips 
corresponding to the apparition of new Dirac cones. The energy position of these 
dips in the calculated DOS presented in Fig.~\ref{doszzarm} using $U = 1.04$ eV 
(red dots for ASL and blue line for ZSL) is in agreement with the following 
analytical formula (vertical dashed lines) given by Park \textit{et 
al.}~[\onlinecite{Park2}] up to $E=1$ eV , which confirms that the position of 
the new Dirac cones only depends on $L$:
\begin{equation}
E_m=\pm \hbar v_F \frac{m \pi}{L}
\label{ParkFormula}
\end{equation}
with $v_F$ the Fermi velocity in pristine graphene and $m$ an integer. At higher 
energies, the agreement between numerics and analytics becomes gradually worse, 
but this is not relevant because the simplified first neighbor TB model loses 
its pertinence there and because such energies are in principle experimentally 
inaccessible to electronic transport measurements. 

In the DOS, each dip associated to a Dirac cone is preceded by a peak due to 
secondary Van Hove singularities (VHS) in the electronic band structure. The 
amplitude of these VHS is influenced by the potential $U$. 
For small values of $U$, the dips and associated secondary VHS are not visible 
in the DOS although Dirac cones can be identified in the electronic band 
structure (not shown here). 

For values of $W$ different from the symmetric case (\textit{i.e.} $W \neq 
L/2$), an electron-hole asymmetry in the DOS is created, and the amplitudes of 
the secondary VHS change. Nevertheless, the energy position of the peaks is 
robust to this change of $W$. For a given $L$, the DOS curves corresponding to 
SLs given by widths $W$ and $|L/2-W|$ are antisymmetric to each other around the 
CNP [see panels (b) and (c) of Fig.~\ref{doszzarm}]      

\begin{figure}[h!]
\center 
\includegraphics[width=9cm]{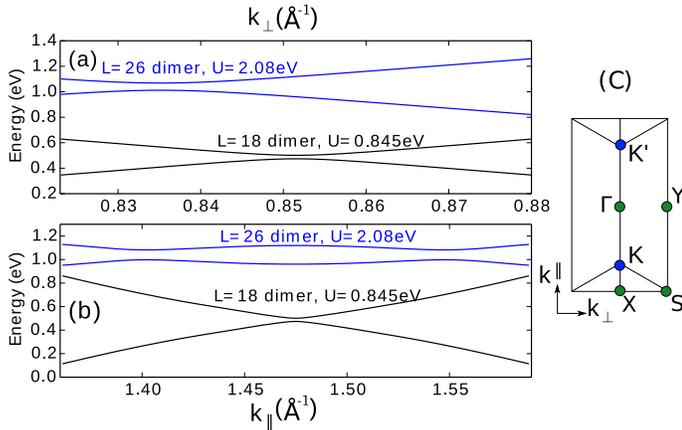}
\caption{Band structure close to the $K$ point for two ASLs ($U = 0.845$ and $U 
= 2.08$, depicted in black and blue, respectively) with band gap in the 
$k_\perp$(a) and $k_\parallel$ (b)  directions. The CNP was not realigned at 
$E=0$ for this figure. As expected, its position is given by $U/2$ (because 
$L=W/2$ in Eq.~\ref{shiftCNP}). (c) Brillouin zone (BZ) associated to the unit 
cell, the position of the Dirac cone (if it exists) is labeled $K$ in reference 
to its position in the original hexagonal BZ of graphene. For ASLs, if multiples 
cones do not appear, the conduction and valance band still present multiple 
maxima and minima in the $k_\parallel$ direction when $L$ and $U$ are large 
enough [e.g. (b), blue case].}
\label{gap1}
\end{figure}

\subsection{Superlattice barrier edge-dependent characteristics}
\label{DOSEdge}

\begin{figure}[htb]
\center
\includegraphics[width=7cm]{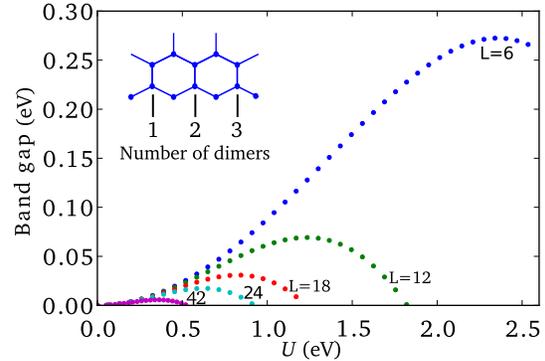}
\caption{Variation of the gap with $U$ for different ASLs. The value of $W$ 
($=L/2$) is chosen so that all the SLs are from the $3p$ family ($n=0$). The 
value of L is given in number of dimers as described by the inset. } 
\label{gap}
\end{figure}

\begin{figure}[tb]
\center 
\includegraphics[width=7cm]{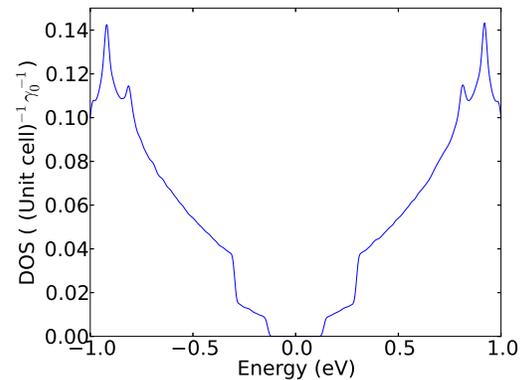}
\caption{Density of state for the ASLs having the largest gap in figure 
\ref{gap} (\textit{i.e.} $L$ containing $6$ dimers and $U=2.34eV$). The gap is clearly 
visible and has the same value as the one obtained directly from de band 
structure, confirming the existence of the observed gaps.} 
\label{gapDOS}
\end{figure}

For sufficiently large values of $U$ and small values of $L$, a splitting of the 
secondary VHS occurs for ZSLs. This splitting is not observed for ASLs, as 
depicted in Fig.~\ref{doszzarm}(a).
In addition, for ASLs, band gaps can appear under certain conditions (see 
Figs.~\ref{gap1}, \ref{gap} and \ref{gapDOS}). Their existence depends on the 
number of carbon dimers $3p+n$ (with $n=0,1$ or $2$ and $p$ an integer) in the 
barrier which allows to group the ASLs into three families defined by the value 
of $n$ (see inset of Fig.~\ref{gap} for visual representation of the dimers). At 
low potential (for $U\gtrsim0$), a gap only opens up for the family where $n=0$. 
Higher values of potential ($U \gg 0$) are required to observe gaps for the two other 
families ($n=1$,$2$). The values of the gaps are generally very small (few meV) as illustrated in Fig.~\ref{gap1}, 
and are inversely proportional to $W$. The exact value of the gap can then be 
fine-tuned by varying $U$ following the bell shape as pictured in 
Fig.~\ref{gap}. The DOS corresponding to the largest calculated band gap is 
shown in Fig.~\ref{gapDOS}.

\section{Velocity at the charge neutrality point}
\label{velocityAtCNP}

\begin{figure*}[tb]
\center
\includegraphics[width=0.8\textwidth, trim =0cm 0cm 0cm 0cm, 
clip]{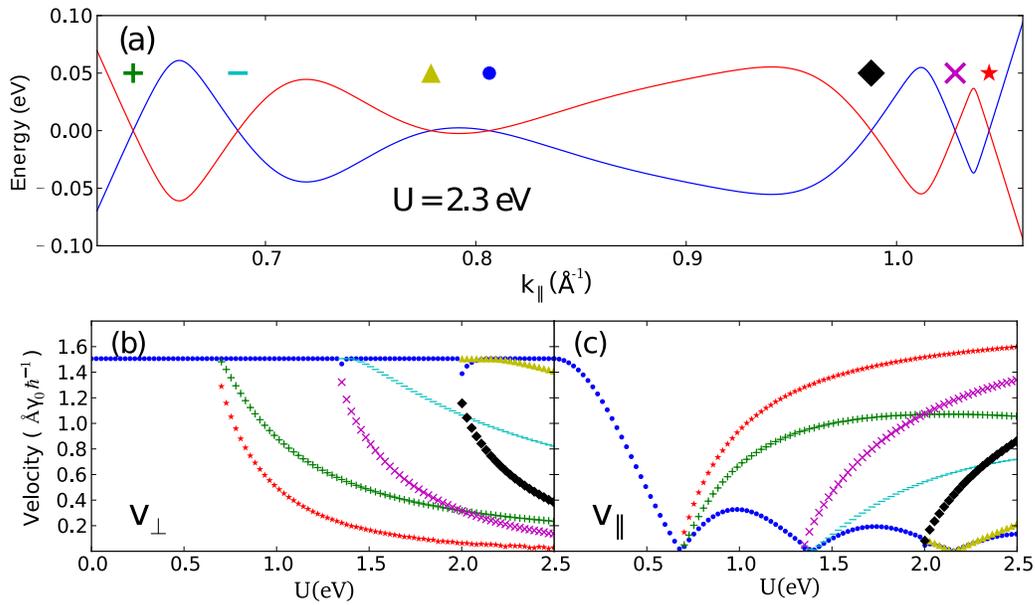} 
\caption{Variation of the velocity at the CNP in the directions perpendicular 
($\perp$) (b) and  parallel ($\parallel$) (c) to the potential barrier, for 
ZSLs.  Each symbol in (b) and (c) is associated to a particular cone as depicted 
in panel (a) representing the band structure for a chosen potential, $U=2.3$ eV, 
in the direction $k_{\parallel}$. $L=10$ nm and $W=L/2$. The \textit{main} (or 
original) cone is always at the center and its velocity is depicted by the blue 
circles. The new cones appear two by two, one on each side of the \textit{main} 
cone. The last cones to appear are therefore closest to the \textit{main} cone. 
Corresponding symbols between (a), (b) and (c) panels are used to facilitate 
reading.}
\label{velV}
\end{figure*} 

Focusing on the central dip in the DOS, \textit{i.e.} at the CNP, the multiplication of cones 
at this energy observed by Ho \textit{et al.}~[\onlinecite{Ho09}] only occurs 
for ZSLs for which $W=L/2$. 
The present simulations agree with the creation-by-pairs model developed by 
Barbier \textit{et al.}~[\onlinecite{Barbier10}]. Nevertheless, our simulations 
indicate that these new cones can be classified into two categories, each with 
different properties from the original cone (labeled \textit{main} Dirac cone 
in the rest of this article). If $W\neq L/2$, these new cones are shifted in 
energy away from the CNP (positioned at zero energy by convention). This 
behavior is discussed in section~\ref{velocityAtEnergies} focusing on the 
energy dependency of the velocity. Large variations of $W$ from the symmetric 
case ($W=L/2$) reduce the number of cones. In other words, if $W$ tends to $L$ 
or $0$ all the new cones disappear. The present section focuses solely on the 
symmetric configuration ($W=L/2$) where all the new cones appearing in ZSLs are 
at the CNP. 

\subsection{Symmetric ZSLs}

As predicted by Park \textit{et al.}~[\onlinecite{Park}], SLs induce an 
anisotropic velocity renormalization. This picture is depicted and extended in 
Fig.~\ref{velV}. The velocities are described as a function of $U$. Because the 
determining factor for this velocity renormalization is actually the product $L 
U$, similar curves (not shown here) can be obtained by varying $L$ and keeping 
$U$ constant. This comment is valid for the remaining of the section.

In Fig.~\ref{velV}, the \textit{main} Dirac cone at the CNP (blue circle symbol) 
has the same flavor as the one described by Park \textit{et 
al.}~[\onlinecite{Park}], namely that the velocity perpendicular to the barrier, 
$v_\perp^m$ ($m$ for \textit{main}) in panel (a), is constant and equals the 
velocity in pristine graphene , while the velocity parallel to the barrier, 
$v_\parallel^m$ in panel (b), varies periodically and goes to zero for certain 
values of $U$. New cones appear at the CNP for symmetric barriers. All the cones 
generated on the \textit{left} side of the \textit{main} cone in panel (c) 
(plus, minus and triangle symbols) are part of a second flavor. The third flavor 
contains the cones created on the \textit{right} side (diamond, cross and star 
symbols) of the \textit{main} cone.

Apart from the two first new cones (plus and star symbols), each set of two new 
cones appears slightly before the minimum of $v_\parallel^m$ (see, for instance, 
diamond and triangle symbols in Fig.~\ref{velV}(c)). The energy difference between this 
minimum and the energy at which the new cones appear increases with $U$ or $L$.

Cones of second and third flavor roughly depict a similar behavior for the 
velocity renormalization with $U$ (or $L$). Indeed, $v_\parallel$ always starts 
from zero and slowly saturates with $U$ ($L$). Surprisingly, the velocity for 
the last cone (star symbols) is higher than the velocity for pristine graphene.
The reason for it contrasts with the squeezing of the Dirac cone due to electron-electron 
interactions \cite{PhysRevLett.109.116802}. A difference in behavior between the 
\textit{left} and \textit{right} cones is also visible: the velocity of the 
\textit{left} cones saturates more quickly than the velocity of the 
\textit{right} cones. 

In the perpendicular direction, all characteristics are inverted: the velocity 
goes from the velocity of pristine graphene toward zero and decreases more 
quickly for the cones on the \textit{right}. Barber \textit{et 
al.}~[\onlinecite{Barbier10}] found a similar behavior for the additional cones. 
Nevertheless this separation into two classes, with slightly different 
properties, was missing in their analysis. They also found that the new cones 
appear at a minimum of $v_\parallel^m$, a conclusion which is slightly modified here.    

For the velocities calculated at directions in between the perpendicular and the 
parallel one, the velocities $v_\perp$ and $v_\parallel$ always appear for all 
families as extrema and the velocity changes smoothly between them.

\begin{figure}[h]
\center
\includegraphics[width=8cm]{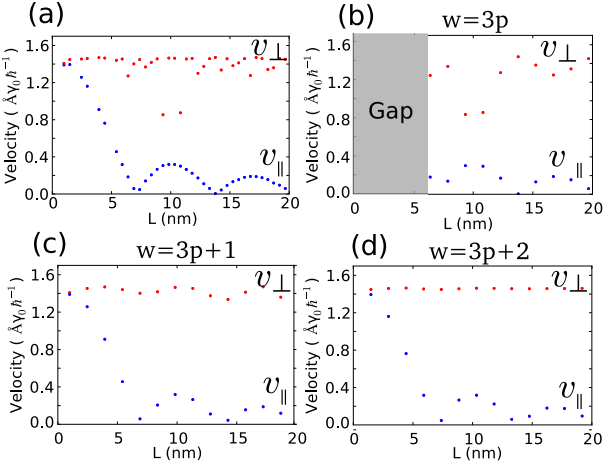}
\caption{Variation of the velocity with $L$ at the CNP for symmetric ASLs for the three 
families : $3p$, $3p+1$ and $3p+2$, with $W=L/2$ and $U=1eV$.  (a) The 
datapoints obtained without distinction between the families of ASLs depict 
non-monotonous variations in the velocity in the $\perp$ direction. Separating 
into the three families (b), (c) and (d), the curves make more sense for this 
direction. In the $\parallel$ direction, the velocity is completely similar to 
what is observed with ZSLs.}
\label{arm-L-m}
\end{figure}

\subsection{Symmetric ASLs}

For ASLs, the impact of the parameters $L$ and $U$ is not completely captured by 
the product $L U$ because of the existence of three families. When $U$ does not 
allow to switch between those families, the value of $L$ can, and this will thus 
induce a different behavior. 

\begin{figure}[tb]
\center
\includegraphics[width=8cm]{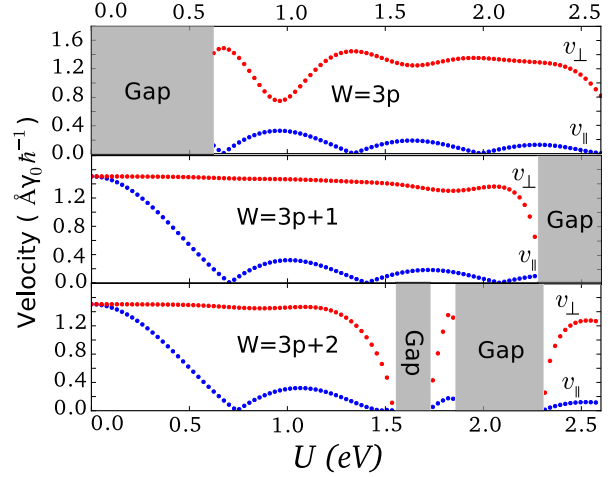}
\caption{Variation of the velocity with $U$ at the CNP for symmetric ASLs for the three 
families: $3p$, $3p+1$ and $3p+2$. The period $L$ used is 21, 19 and 20 dimers 
respectively. The shaded area highlights the position of the gaps. Only the $3p$ 
family has a gap at low potential. Near a gap, the oscillations of $v_{\perp}$ 
increase up to a point where the velocity falls rapidly to zero.}
\label{arm-O-m}
\end{figure}

Figure~\ref{arm-L-m} shows the velocity renormalization for the ASLs case (for 
$W=L/2$) when changing the value of $L$ and figure \ref{arm-O-m} when the value of $U$ is
modified. 
For the renormalization in function of $L$, all data points are gathered in Fig.~\ref{arm-L-m} (a) 
but also separated into 
the three families $n=0$, $1$ or $2$ in the panels (b),(c) and (d), according to 
the previous discussion in Section~\ref{DOSEdge}. After separation into three families, the similarities between variation of $U$ and $L$ are recover [compare Fig.~\ref{arm-L-m}(b),(c),(d) and Fig.~\ref{arm-O-m}(a),(b),(c) respectively].
The overall behavior of $v_\perp$ is similar as the one of ZSLs, provided the 
values of $U$ or $L$ which create a band gap are excluded (inducing an absence of 
velocity at the CNP). 
Since Dirac cone multiplication does not occur for this barrier edge geometry, only the 
velocity of the main Dirac cone is depicted for ASLs. 
For $v_\parallel$, the behavior is similar to ZSLs 
[compare for instance blue circle symbols in Fig.~\ref{velV}(c) and Fig.~\ref{arm-L-m}(a) (or Fig.\ref{arm-O-m})], 
which confirms the $L$ and $U$-dependency highlighted above. 
For $v_\perp$, the behavior is different than for ZSLs. 
Indeed, this velocity now varies, in contrast to the 
constant value observed for ZSLs [Fig.~\ref{velV}(b)]. As displayed in \ref{arm-L-m},
 these variations depend on the family to which the ASLs belong to ($n=0,1$ or $2$), which explains the 
non-monotonous variations in the velocity observed. When approaching the gaps, oscillations of the otherwise constant value of $v_{\perp}$ are observed. Perpendicular velocity eventually vanishes when reaching the gap region (see Fig.\ref{arm-O-m}).

\begin{figure*}[bht]
\center
\includegraphics[width=\textwidth,trim =0cm 0cm 0cm 0cm, clip]{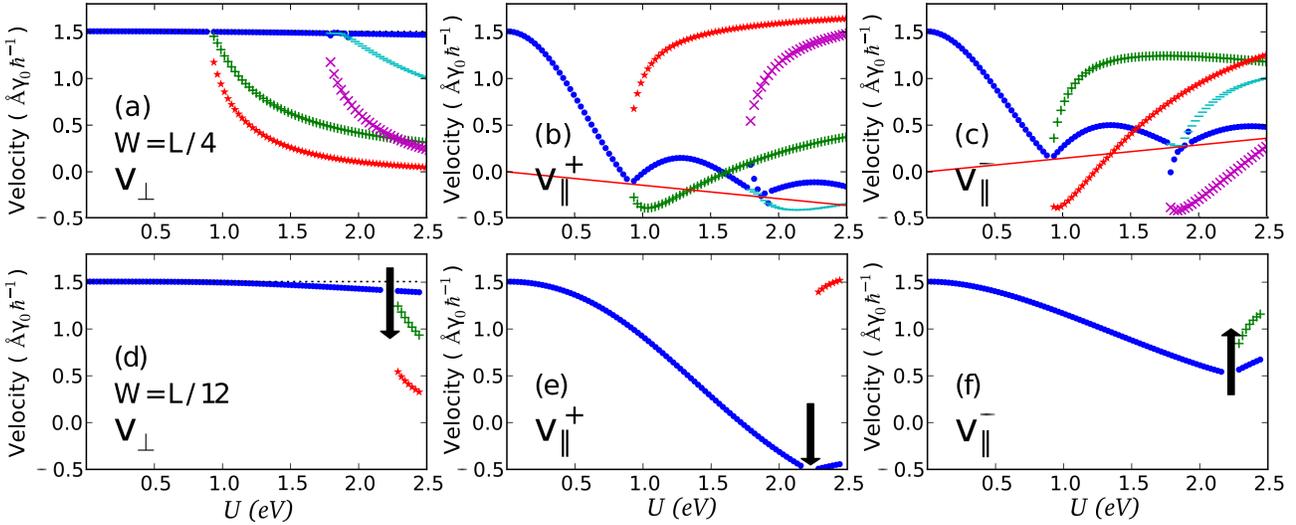} 
\caption{Variations of the velocity with U for asymmetric ZSLs, $L=48$ dimers 
($\sim 10$ nm). The correspondence of the symbols is the same as the one used in 
Fig.~\ref{velV}. (a), (b) and (c): $W=L/4=12$ dimers. (d), (e) and (f): 
$W=L/12=4$ dimers. When $W$ goes towards $0$ ($L$), the asymmetry between the 
$v_\parallel^-$ and $v_\parallel^+$ increases. The black arrows in (d), (e), and 
(f) indicate the direction of the curve when varying $W$ towards zero. The 
reduction is non-linear and increases when W is closest to $0$. Similarly, the 
$U$-dependency of the curves is also dilated when $W$ decreases (transition 
between upper and lower panels). At the limit $W=0$, the dilation would become 
infinite, recovering the properties of pristine graphene.}

\label{valvarWe2}
\end{figure*}

\subsection{Asymmetric ZSLs}

An asymmetry between the modified and pristine zone ($W \neq L/2$) has also a 
large impact on the velocity renormalization. The effect at the CNP is larger 
for ZSLs. Because of the asymmetry, the new Dirac cones are not generated at the 
CNP anymore (and those will thus be discussed in the next Section). In this paragraph, only 
the impact on the \textit{main} cone (blue circle symbols in Fig.~\ref{valvarWe2}) is 
considered.
  
The first effect of this asymmetry in ZSLs is to break the equivalence between 
the $k_{\parallel}^+$ and $k_{\parallel}^-$ direction for $v_\parallel$ (see  
Fig.\ref{conePosband} for sign convention).
This difference in the behavior of velocities  $v^+_{\parallel}$ and 
$-v^-_{\parallel}$ is illustrated in Fig.~\ref{valvarWe2}.

The blue circles in panels Fig.~\ref{valvarWe2}(b) and (c) suggest that, for a 
given valley K, some asymmetry is obtained depending on the direction of the 
carriers flow through the barrier for $U\geq1$ eV. Nevertheless, a 
correspondence exists between the two velocities. The velocity 
$v^+_{\parallel}$($-v^-_{\parallel}$) obtained for a width $W$ is exactly the 
same as the velocity $-v^-_{\parallel}$ ($v^+_{\parallel}$) obtained for a width 
$L-W$, respectively.

Looking at the impact of $U$, both $v_{\parallel}$ and $v_{\perp}$ are affected, 
in opposition to the symmetric case. $v_{\perp}$ now presents a decreasing 
behavior with $U$, which gets more pronounced when $W$ tends towards $L$ or $0$ 
and completely disappears at $W=L/2$, consistent with previous observations.    
 
In Fig.~\ref{valvarWe2}(a), for $W=L/4$, the decrease is not very pronounced in 
comparison with the (constant) dotted line found for $W=L/2$. As $W$ decreases 
further, the renormalization of $v_{\perp}$ gets more pronounced [see panel 
(d)]. 
The effect of $U$ has a larger impact on $v_{\parallel}$ [panels (b) and (c)]. 
First, the minima of $v_{\parallel}$ are not any more at zero. Then, the behavior of 
$v^-_{\parallel}$ and $v^+_{\parallel}$ is opposite: if the value of the minima 
increases for $-v^-_{\parallel}$ [as in panels (c) and (f)], the value of the 
minima in $v^+_{\parallel}$ decreases [as in panels (b) and (e)]. In a first 
approximation, if $W$ is not too small, the minima are aligned on the red line 
starting from zero [panels (b) and (c)]. The slope of this line is opposite for 
$v^+_{\parallel}$ and $-v^-_{\parallel}$ and decreases when $W$ goes to $L/2$, 
eventually recovering the behavior of the symmetric barriers, where both 
velocity directions are degenerated. However, if $W$ is small (a few percent of 
the value of $L$), the line joining the minima does not cross the origin of the 
plot [\textit{i.e.} (0,0)] anymore, and is different for both directions (not 
shown here). In such case, the larger the velocity renormalization is in one 
direction (take for instance $v^-_{\parallel}$), the smaller it is in the 
opposite direction ($v^+_{\parallel}$), as already visible by comparing panels 
(e) and (f). 

\begin{figure*}[bht]
\center
\includegraphics[width=\textwidth]{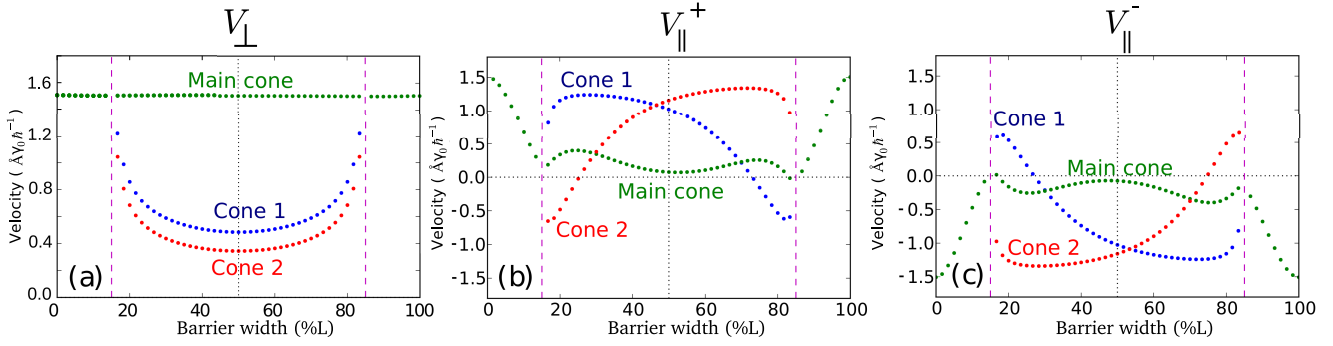} 
\caption{\small Variations of the velocity with the barrier width when several 
cones are present for ZSLs. $L=120$ dimers ($\sim 26$ nm). A large value of $L$ 
is used to have more points in the curve. Indeed, an integer number of dimers 
for $W$ and $L$ is required. Therefore, the larger the value of $L$, the larger 
the number of points that can be computed. $U=0.52$ eV. (a) The velocity in the 
$\perp$ direction is constant for the \textit{main} cone and varies from the 
velocity of pristine graphene to a minimum situated at $W=L/2$ for the other 
cones. (b) and (c) represent the values of the velocity in the two inequivalent 
branches of the cones for the parallel direction. Additional cones are only 
found in the central zone with a L between $\sim$16\% and $\sim$84\%. The 
variations of velocity of two new cones are opposite. Contrarily to the main 
cone, an inversion of the velocity is observed for the new cones. Results 
obtained for smaller value of $L$ are similar.}
\label{Lmulticone}
\end{figure*}

Finally, by varying $W$ (see the green dots corresponding to the main cone 
in Fig.~\ref{Lmulticone}), the impact of asymmetric barriers becomes apparent as 
well. In this situation, Figs.~\ref{Lmulticone}(b) and (c) can be obtained 
from one another by central symmetry around the point $(v=0, W=L/2)$ (as already 
mentioned at the beginning of this paragraph). The other curves in this Fig.  
(blue and red dots in Fig.~\ref{Lmulticone}) are left for 
Section~\ref{velocityAtEnergies}, where the velocities away from the CNP are 
discussed.

\subsection{Asymmetric ASLs}

\begin{figure}[tb]
\center
\includegraphics[width=0.4\textwidth,trim =0cm 0cm 0cm 0cm, 
clip]{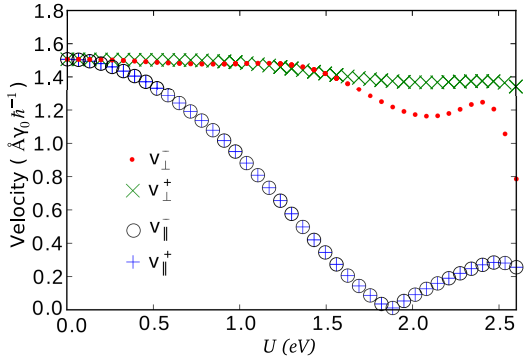} 
\caption{Variations of the velocity with $U$ for asymmetric ASLs. $L=40$ dimers 
($\sim 10$ nm) and $W=L/10=4$ (3p+1 family). No anisotropy exists for 
$v_\parallel^-$. In opposition to ZSLs, $v_{\perp}$ exhibits an anisotropy 
between the $+$ and $-$ directions.}
\label{valvarW-ASL}
\end{figure}

For ASLs, the asymmetry induced by the value of $W$, has a weaker influence on 
the velocity renormalization, in comparison with ZSLs. The values of 
$v_{\parallel}$ stay symmetric (\textit{i.e.} no degeneracy lifting between 
$v^+_{\parallel}$ and $v^-_{\parallel}$) for $W \leq L/2$ and the dependency on 
the $U$ parameter is similar to the one observed for $W=L/2$. In contrast, the 
degeneracy is lifted for $v_{\perp}$, albeit quite softly. More specifically, 
the oscillations in the vicinity of the band gaps are different, as depicted in 
Fig.~\ref{valvarW-ASL} for the $3p+1$ family (for the position of the band gaps 
for different families, see Fig.~\ref{arm-O-m}). 

\section{Energy dependency of the velocity}
\label{velocityAtEnergies}

As mentioned previously, the periodic potential does not only modify the 
velocity at the CNP, but also impacts the velocity at higher energies. Those 
energies are the topic of this Section. On the one hand 
(Section~\ref{smallEnergy}), for energies close to the CNP ($\leq 0.1 eV$), only 
the asymmetric ZSLs ($W \neq L/2$) are discussed, for which  the new cones shift 
away from the CNP. ASLs do not induce additional cones, and are thus excluded 
from this discussion. On the other hand (Section~\ref{largeEnergy}), at higher 
energies ($\gg 0.1 eV$), the differences between ASLs and ZSLs disappear. 
The focus is thus shifted towards the velocities at intermediate incident angles.

\subsection{Asymmetric ZSLs}
\label{smallEnergy}

Going back to Fig.~\ref{valvarWe2}, showing the impact of the asymmetry on the 
additional cones (plus, minus, cross, and star symbols)
as a function of U, the inequivalence between $k_\parallel^+$ and $k_\parallel^-$ directions 
is clearly apparent, as it was already observed for the \textit{main} cone (blue circles).

\begin{figure*}[tb]
\center
\includegraphics[width=\textwidth]{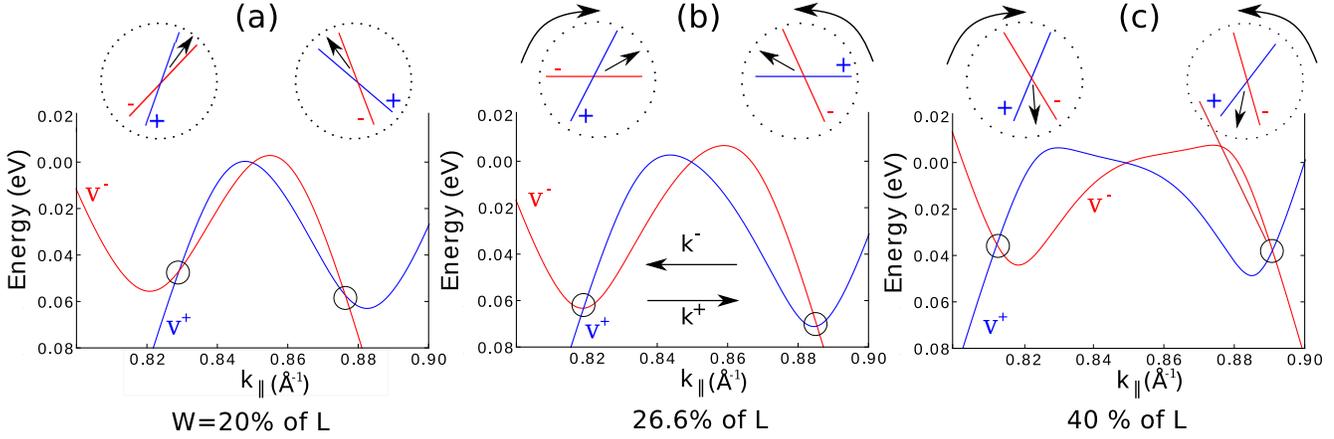} 
\caption{Rotation of the Dirac cones with the modification of $W$ for ZSLs. The 
arrows outside the circle above the graph indicate the fictive rotation 
direction when transitioning the values of $W$ from (a) to (c). The arrow inside 
the cone has no physical meaning, but helps visualize the cone rotation.. $v^+$ 
and $v^-$ are defined as the velocity on the two branches of the cones. The 
corresponding branches are color-coded in the mainframes below. $v^-$ is taken 
on the branch showing an overall decreasing behavior and $v^+$ an overall 
increasing behavior. (a) Both branches of the cone have the same slope: $v^-$ 
and $v^+$ have the same sign. (b) One of the bands is flat: $v^-$ or $v^+$ is 
zero. (c) The two branches have opposite slopes but the cone is still slightly 
turned: $v^-$ and $v^+$ have opposite signs but different values.}
\label{conePosband}
\end{figure*}

Varying now the parameter $W$ (Fig.~\ref{Lmulticone}), two zones 
can be distinguished (separated by dashed vertical purple lines). In the first 
zone (outer region, where $W$ is close to $0$ or $L$), the value of $|L-W|$ is 
too small to create multiple cones (only the main cone 
exists). A simple renormalization of the velocity parallel to the barrier ($v^-_{\parallel}$), 
similar to the one by changing the $U$ parameter, is observed. 
In the second zone (central region), the velocities corresponding to 
the additional cones show a more exotic behavior. In particular, 
close to the boundary between the two zones, the velocities of the 
additional cones have the same sign in $v^+_{\parallel}$ and $v^-_{\parallel}$. 
This is somehow unusual because it means that the slope of the two bands forming the cone in 
the $\parallel$ direction have the same sign. This curious behavior of the 
additional cones can be visualized by looking at the evolution of the band 
structure as a function of $W$ in Fig.~\ref{conePosband}. For $W=L/5$ [panel (a)], 
the cones are tilted in such a way that the velocities have the same sign 
(upper insets zoom in on this peculiar behavior). For higher values of $W/L$ 
(here at $26.6\%$), one of the branches becomes parallel to the $k_{\parallel}$ 
axis and thus the velocity drops to zero [see panel (b)]. After this transition 
value, a situation occurs where the velocities in the cone have opposite signs 
[see panel (c)]. The absolute values of these velocities are not yet 
equal because the cone is still slightly rotated. By increasing further the 
value of $W$, the velocities become closer and closer to each other in absolute 
value, to finally recover the symmetric barriers case discussed in previous 
Sections.

\subsection{Angle dependency}
\label{largeEnergy}

\begin{figure}[tb]
\center
\includegraphics[width=8cm]{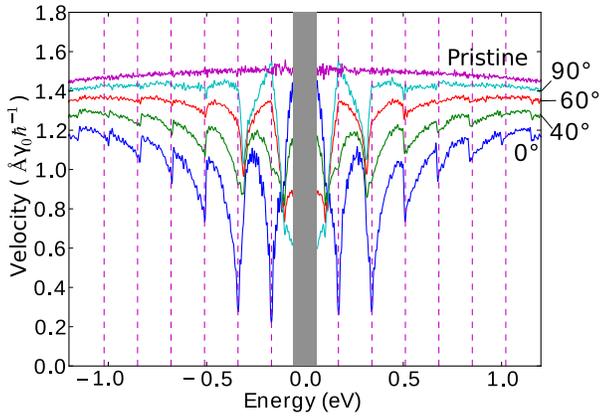} 
\caption{\small Energy dependency of the velocity at different angles, for ZSLs. 
The first curve on the top is for pristine graphene. The direction perpendicular 
to barrier is taken as $0^\circ$ reference. From top to bottom the angles are 
$90^\circ$ ($v_{\parallel}$), $60^\circ$, $40^\circ$ and $0^\circ$° 
($v_{\perp}$) . The vertical dashed lines represent the position of the new 
Dirac points generated by the SLs as given by Eq.~\ref{ParkFormula}. The 
potential applied is $U=0.52$eV with a period of $L=10$nm ($W=L/2$). The area 
around the CNP is shaded because the results are numerically too unstable 
there.}
\label{fAVE}
\end{figure}

Up to now, only the $0^\circ$ ($\perp$) and the $90^\circ$ ($\parallel$) cases 
have been discussed, being the most straightforward. We found that the velocity 
changes smoothly between these two extreme cases. The present Section extends 
the analysis for these intermediate angles at energies away from the CNP, 
leading to richer physics. For this analysis the velocities are computed
within the Kubo-Greenwood method in the ballistic regime.
ZSLs and ASLs behave very similarly at higher 
energies. Therefore, only ZSLs are depicted. 

In Fig.~\ref{fAVE}, both the angle and the energy dependency are clearly 
visible. At the CNP, the evaluation of velocity is hindered by numerical divergences
(shaded region in Fig.~\ref{fAVE}), difficult to resolve using the diagonalization trick, because of the angle dependency.
For energies very close to the CNP, the low-angle velocity (close to 
$0^\circ \equiv v_{\perp}$) is larger than the high-angle velocity (close to 
$90^\circ \equiv v_{\parallel}$). For higher energies, \textit{i.e.} away from the CNP, the 
behavior is inverted. The transition between these two regimes occurs around 
$0.12$ eV in Fig.~\ref{fAVE}, which corresponds to the position of the first 
peak (VHS) in the DOS.

For even higher energies, periodic oscillations in the velocity appear. 
The amplitude of these oscillations is maximum for smallest angles ($v_{\perp}$) and become barely 
noticeable for largest angles. Every minimum of the velocity corresponds to the 
position of a Dirac cone in the electronic band structure, which corresponds 
also to a minimum in the DOS. Even if the minima are not apparent in the DOS 
(see Fig.~\ref{doszzarm}), they are clearly visible for the velocity up to 
approximately $1$ eV in Fig.~\ref{fAVE} for low angles. 

As already observed throughout this paper, a minimum in the DOS caused by a 
Dirac point does not necessarily imply a maximum in the velocity. This is 
further confirmed with this energy dependent renormalized velocity curve. 
Changing from $90^\circ$ to $0^\circ$, a maximum in the velocity can become a 
minimum at the position of the additional Dirac cones. In other words, the Dirac 
cones away from the CNP induce a local maximum of the velocity in the direction 
parallel to the barrier and a minimum in the direction perpendicular to the 
barrier. Finally, the energy-dependent oscillations in the velocity are 
more pronounced with increasing values of $U$ (not shown here).

\section{Effect of disorder}
\label{disorderEffectSection}

All the potential barriers considered so far were ideal and displayed a perfect periodicity,
keeping therefore the charge carriers in the ballistic regime.
The absence of random disorder precludes quantum interference phenomena, such as weak and 
strong localization effects. 
Therefore, to observe the transport signatures from this paper, experiments should aim at 
minimizing any form of extrinsic disorder (ad-atoms, vacancies, trapped screened 
charged impurities) that may lead to strong scattering. 
In addition, the barriers themselves should be free of disorder and atomically perfect. 
This situation is obviously rather far from real experimental conditions.
Existing literature gives guidance on which features should remain robust and which features might 
disappear if such disorder becomes too strong. More specifically, for 
uncorrelated white-noise disorder on the barriers, a perpendicular incident 
angle keeps a robust transmission~\cite{abedpour2009conductance}, suggesting 
associated transport features to be reliable, while transmission is strongly 
reduced when increasing the incident angle. The system should therefore be kept 
as clean as possible to observe parallel conductivity features (the 
perpendicular case being less sensitive). 
To counter the detrimental effect of white-noise disorder on the transmission, 
a long-range correlation between the potential barriers may be used~\cite{barbier2010single,esmailpour2012conductance}.\\ 

\begin{figure}[tb]
\center
\includegraphics[width=0.95\columnwidth]{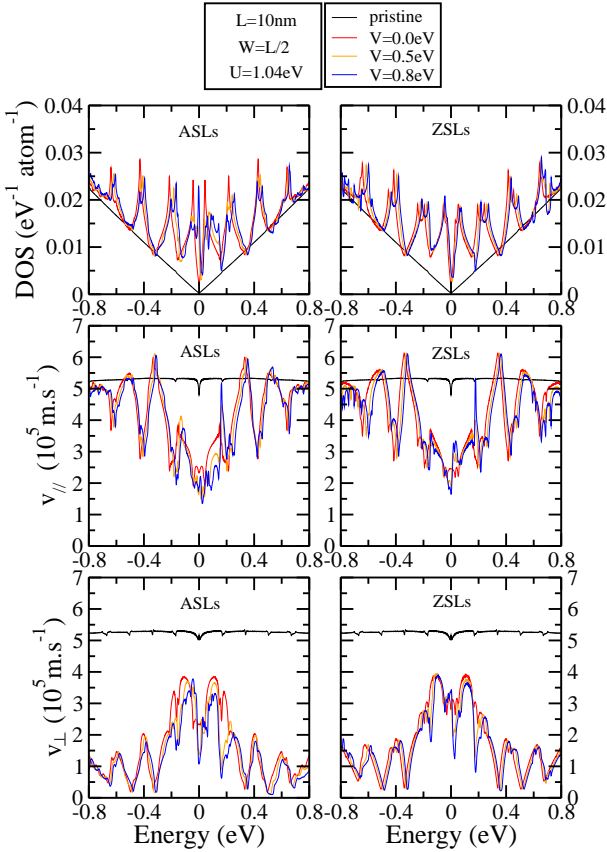} 
\caption{\small Density of states and velocities for ASLs and ZSLs ($L$=$10$ nm, $W$=$L/2$, $U$=$1.04$ eV) 
without ($V$=$0$ eV) and with Anderson disorder ($V$=$0.5;0.8$ eV).}
\label{disorder}
\end{figure}

The systematic study of the effect of disorder on electronic transport in 
graphene 1D SLs is out of the scope of this article.
However, selected prospective simulations of disordered graphene 1D SLs are presented in Fig.~\ref{disorder}.
The upper panels compare the effect of Anderson (white-noise) disorder 
on the DOS of both ASLs (left) and ZSLs (right). Anderson disorder is introduced 
by randomly varying the onsite potentials of all atomic sites with a value of $\delta \varepsilon_{p_z} \in [-V/2 , +V/2]$. The existence and the position of the new Dirac points remain noticeably robust, 
up to $V = 0.8$ eV. Equation~(\ref{ParkFormula}) can still be used to estimate their position. 
However, disorder induces a splitting in the secondary VHS for the ASLs, 
while this splitting already exists in absence of disorder for ZSLs. Smoothed curves and VHS peaks may be expected by averaging over a manifold of disorder configurations. This makes it very difficult to experimentally differentiate between ASLs and ZSLs, based on the DOS.In Fig.14, both parallel (central panels) and perpendicular (lower panels) velocities are plotted for ASLs (left panels) and ZSLs (right panels).The black curves give the velocity for pristine graphene, 
in absence of SLs and Anderson disorder. Applying the Anderson disorder (yellow and blue curves) 
on top of the clean SL (red curve) does not change much the behavior of both velocity directions above $\pm 0.3$ eV, 
which is as expected due to the selected $V$ values. 
Nevertheless, the qualitative features (maxima and minima) remain globally robust 
even for low energies. At zero energy, a drop by a factor two for $V=0.8$ eV is observed 
for both velocities for the ASL case. For the ZSLs case, the changes in velocity at low energy are smaller, 
with the perpendicular features even more robust, in agreement with above statements based on literature. 
Nevertheless, for both SL orientations, a large drop (increase) of the perpendicular (parallel) velocity 
appears around $0.2$ eV, respectively.
Further calculations would be required to complete this picture, 
but these preliminary velocity and DOS calculations indicate that 
ZSLs are more robust to the detrimental effect of Anderson disorder than ASLs.

\section{Conclusion}

The impact of the 1D SL orientation with respect to the graphene crystal was examined on grahene 1D SLs physics were studied. Important differences were 
highlighted, like the presence of new cones at the CNP for zigzag SLs and the 
opening of gaps for armchair SLs. A specific effect induced by the SL alignment, 
absent in the literature, was found in the velocity renormalization for the 
direction perpendicular to the 1D potential. This renormalization occurs in 
ASLs, in general, and in ZSLs when the barriers are asymmetric. On the other 
hand,  the velocity in the direction parallel to the 1D potential behaves 
similarly to what is predicted in the literature, although the position of the 
minima can be modulated by the type of SLs or the parameters used. The asymmetry 
of the SLs was shown to have a strong impact on the velocity. In particular, 
it can break the initial symmetry between the forward and backward momentum 
direction with respect to the Dirac cone symmetry for the velocity in the 
perpendicular direction for ASLs and parallel direction for ZSLs. This breaking 
of the symmetry can be interpreted by a rotation and deformation of the Dirac 
cone(s), leading to strong modifications in the associated velocities. By 
studying the angle dependency of the velocity through the barrier, the smooth 
transition between the parallel and perpendicular direction is understood. 
The calculated gaps in ASLs are very small. More advanced theoretical 
frameworks, such as \textit{ab initio} simulations, are required to correctly 
assess their amplitude. Further studies may focus on systems with mixed chiral orientation different than pure armchair or zigzag orientation reproducing certain experimental conditions. Nevertheless, much improved 
control of edge geometry (using a bottom-up approach for chemical synthesis\cite{doi:10.1021/nl801316d,
sprinkle2010scalable,cai2010atomically,kato2012site}) 
make the present systems promising for electron collimation experiments.

\acknowledgments
 A.L. and J-C.C. acknowledge financial support from the F.R.S.-FNRS of Belgium. 
Computational resources have been provided by the supercomputing facilities of 
the Universit{\'e} catholique
de Louvain (CISM/UCL) and the Consortium des Equipements de Calcul Intensif en 
F{\'e}d{\'e}ration Wallonie Bruxelles (CECI).
This research is directly connected to the ARC on Graphene Nano-electromechanics (N$^{\circ}$ 11/16-037) sponsored by the Communaut\'e Fran\c{c}aise de Belgique and to the European ICT FET Flagship (N$^{\circ}$ 604391) entitled ``Graphene-based revolutions in ICT and beyond''.   

\appendix
\section{Numerical instability at the charge neutrality point}
\label{instability} 

Numerical instabilities make the accurate calculation of the velocity at the CNP difficult. 
In the following we show why and how numerical instabilities appears at the CNP.
This feature is illustrated within the Kubo-Greenwood approach which we further describe here.

The Kubo-Greenwood technique was originally developed to compute the 
conductivity from the quantum mechanics point of 
view~\cite{PhysRevLett.79.2518,PhysRevB.59.2284,PhysRevB.84.235420,Roche20121404} 
and gives information on both the quantum and the semi-classical transport. 
The energy dependent carrier velocities in the ballistic regime can also be 
investigated using this formalism. 
The formalism is based on the propagation of a wavepacket throughout the material, described by the
diffusion coefficient $D$ defined as:
\begin{equation}
D(t)=\frac{\partial }{\partial t}\Delta R^2(t) =\frac{\partial }{\partial t} 
\Delta X^2(t) + \frac{\partial }{\partial t} \Delta Y^2(t)
\label{diff}
\end{equation}
where $\Delta R^2(t)$ is the mean quadratic spreading of the wavepacket. 
$\Delta X^2(t)$ and  $\Delta Y^2(t)$ are the mean quadratic spreadings in the $x$ and $y$ direction, respectively.
The behavior of D with time indicates the transport 
regime in which the wavepacket resides~\cite{thesisaurelien}. 
If the diffusion coefficient increases linearly with time, 
the electrons are not scattered and move freely in the material (ballistic regime). 
When this coefficient saturates to a certain value ($D^\text{max}$), the 
electrons have experienced sufficient scattering to reach the diffusive regime. 
A further increase or decrease of the coefficient indicates the onset of
(anti-)localization~\cite{Roche20121404,PhysRevB.84.235420}, rooting in quantum localization corrections. 
The diffusion coefficient $D$ depends on the mean quadratic spreading of the 
wavepacket as described in Eq.~(\ref{diff}), while the spreading in a given direction (say, $x$) is given by:
\begin{widetext}
\begin{equation}
\Delta X^2(t)=\left\langle | \hat{X}(t)-\hat{X}(0) |^2 \right\rangle_E
= \frac{ Tr \left[ \left(\hat{X}(t)-\hat{X}(0)\right)^{*} \delta(E-\hat{H})
 \left(\hat{X}(t)-\hat{X}(0)\right) \right]}
 { Tr \left[ \delta(E-\hat{H}) \right]  }
 \label{xt2}
\end{equation}
\end{widetext}
where the operator $\hat{X}(t)$ is the position operator in the Heisenberg picture. 
The numerator, and the denominator (corresponding to the DOS),
are computed separately, with the same Lanczos algorithm using continued fractions. 
To reduce the computational cost, the trace is replaced by an average of random phase states obtained by adding 
a random phase factor to the wavefunction at each orbital of the system. 
Averaging over about $10$ random phase states is usually enough to reach a 
satisfactory convergence ($<1\%$ of variations in the quantities of interest).  
 
The velocity in the ballistic regime can be extracted from the mean quadratic expansion as  
\begin{equation}
\Delta X^2(t)=v_x^2t^2 \Rightarrow v_x=\frac{\sqrt{\Delta X^2(t)}}{t}
\label{vel}
\end{equation}

In pristine graphene, at the CNP, both the numerator and the denominator of 
$\Delta X^2(t)$ [Eq.~(\ref{xt2})] tend to zero. 
The denominator being the graphene DOS, at the Dirac point this term obviously tends to zero. 
From simple physical considerations using Eq.~\ref{vel}, one can show that the numerator of $\Delta X^2(t)$ also tends to zero.
Indeed, a finite value of velocity implies a finite value of $\Delta X^2(t)$.
Since the denominator of $\Delta X^2(t)$ tends to zero in Eq.~(\ref{xt2}), 
a finite spreading can only exist if the numerator tends to zero too. 
Mathematically, using the concept of limits, this is no problem. 
From a numerical point of view, using floating point arithmetics, 
the division of two very small numbers generates large errors, 
explaining the aforementioned instability. 
A similar problem can arise when calculating the energy dependent velocity with 
a direct diagonalization approach. However in that case, the accuracy is much more controlled 
and the results can be improved by using denser k-point meshes in the Brillouin zone.

The Kubo-Greenwood approach was nevertheless helpful in this article to
compute the energy dependent velocity at any intermediate angles as shown in Fig~\ref{fAVE}.

\bibliography{bib}

\end{document}